\documentclass[prb floatfix, twocolumn, superscriptaddress]{revtex4-2}
\def\be{\begin{equation}}
\def\ee{\end{equation}}
\def\bea{\begin{eqnarray}}
\def\eea{\end{eqnarray}}
\def\bi{\begin{itemize}}
\def\ei{\end{itemize}}
\usepackage{braket}
\usepackage{xcolor}
\usepackage{graphicx}
\usepackage{amsmath,amstext,amssymb,bm,color,times}
\usepackage[english]{babel}
\usepackage[T1]{fontenc}
\usepackage[utf8]{inputenc}
\usepackage[colorlinks=true,citecolor=blue,linkcolor=magenta]{hyperref}

\newcommand{\change}[1]{{\color{blue}{{#1}}}}

\begin{document}

\title{\bf Quantum Dark Soliton: \\
           Non-Perturbative Diffusion of Phase and Position }

\author{ J.~Dziarmaga }

\address{ Instytut Fizyki Uniwersytetu Jagiello\'nskiego,
              Reymonta 4, 30-059 Krak\'ow, Poland  }

\date{ September 15, 2004 }

\begin{abstract}
The dark soliton solution of the Gross-Pitaevskii equation in one
dimension has two parameters that do not change the energy of the
solution: the global phase of the condensate wave function and the
position of the soliton. These degeneracies appear in the Bogoliubov
theory as Bogoliubov modes with zero frequencies and zero norms. These
``zero modes'' cannot be quantized as the usual Bogoliubov quasiparticle
harmonic oscillators. They must be treated in a non-perturbative way. In
this paper I develop non-perturbative theory of zero modes. This theory
provides non-perturbative description of quantum phase diffusion and
quantum diffusion of soliton position. An initially well localized
wave packet for soliton position is predicted to disperse beyond the
width of the soliton.  
\end{abstract}

\maketitle


\section{Introduction}

A dark soliton in a quasi one dimensional (1D) atomic Bose Einstein
condensate (BEC) is a particle-like solution of the Gross-Pitaevskii
equation \cite{GS}. Dark solitons were observed in two 
experiments
\cite{Hannover,snake}. A classical soliton in a quasi-1D condensate in a 
harmonic
trap behaves like a (negative mass) particle in a harmonic potential with
a frequency equal to trap frequency divided by $\sqrt{2}$ \cite{GS}. This
results in a (negative frequency) anomalous mode in the spectrum of the
Bogoliubov theory. This mode describes small fluctuations of the soliton
around the center of the trap. A soliton wave packet which is initially
localized in the center of the trap is going to disperse \cite{greying}.
The width of the wave packet grows until it becomes comparable to the
width of the soliton - the healing length. In Refs.\cite{greying} this
dispersion was estimated to happen, for reasonable experimental
parameters, on a time scale of $10$ms.  This quantum ``instability'',
present even at zero temperature, adds to the list of more classical
decay channels \cite{Gora,Durham}.

The calculations in Ref.\cite{greying} were done within Bogoliubov theory
which is a linearized theory of small quantum fluctuations around a
classical soliton localized in the center of the trap. This theory breaks
down when fluctuations grow large because for large fluctuations one
cannot neglect interactions between Bogoliubov modes. The perturbative
theory breaks down when the width of the soliton wave packet becomes
comparable with the healing length i.e. after around $10$ms from soliton
creation. To extend the soliton diffusion beyond this point one has to
treat the diffusion in a non-perturbative way.

In this paper I develop non-perturbative theory of soliton diffusion that
includes both position and phase fluctuations. To avoid some technical
problems, and also to afford more pedagogical presentation, I consider a
soliton in uniform 1D condensate confined to a box of finite length. The
finite length of the box makes quantum and thermal depletion from the 1D
condensate finite. In fact finite quasi-1D condensates were observed in
experiments \cite{1D,Hannover} in harmonic traps. A quasi-1D condensate in
the Thomas-Fermi regime can be considered locally uniform on lengthscales
comparable to the soliton width. What is more atom chips \cite{chips} make
possible confinement of cold atoms in a quasi-1D box.

\section{ Dark Soliton in a Box }

\begin{figure}
\includegraphics[width=8.5cm]{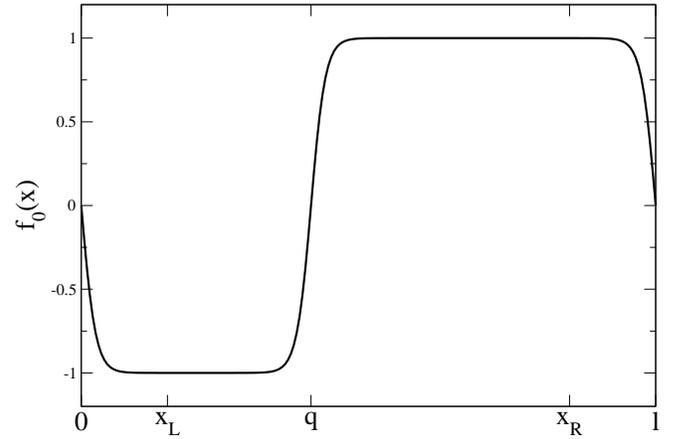}
\caption{
The function $f_0(x)=\frac{\phi_0(x)}{e^{-i\theta}\sqrt{\rho}}$. $x_L$ and
$x_R$ are placed six healing lengths from the left and right walls of
the box.}
\label{figphi0}
\end{figure}

The dark soliton \cite{GS} is a stationary solution of the 
Gross-Pitaevskii equation in 1D 
\begin{equation}
i\hbar\partial_t \phi=
-\frac{\hbar^2}{2m}\partial_x^2 \phi+
g|\phi|^2\phi-\mu\phi.
\label{TDGPE}
\end{equation}
Here $m$ is atomic mass and $g$ is effective 1D interaction strength. The
system is placed in a box by imposing the boundary conditions 
$\phi(x=0)=0$ and $\phi(x=l)$ at the walls of the box. The stationary dark 
soliton is
\begin{equation}
\phi_0(x)=
\left\{
\begin{array}{lcl}
-e^{-i\theta}\sqrt{\rho}\tanh\frac{x}{\xi}  &,& x_L<x     \\
e^{-i\theta}\sqrt{\rho}\tanh\frac{x-q}{\xi} &,& x_L<x<x_R \\
e^{-i\theta}\sqrt{\rho}\tanh\frac{l-x}{\xi} &,& x_R<x
\end{array}
\right.
\label{GS}
\end{equation}
see Figure \ref{figphi0}. Here $\rho$ is (linear) density of the
condensate, $\xi=\hbar/\sqrt{mg\rho}$ is the healing length, $c=\sqrt{\rho
g/m}$ is the speed of sound, and $\theta$ is arbitrary global phase. I
made a convenient choice of $\mu=g\rho$. I assume that the width of the
soliton is much less than the size of the box, $\xi\ll l$, and that the
position of the soliton $q$ is at safe distance of a few healing lengths
from the walls.

This stationary solution is degenerate with respect to the soliton 
position $q$ and to the global phase $\theta$ of the condensate wave 
function.


\section{ Small fluctuations around dark soliton }

The Gross-Pitaevskii equation (\ref{TDGPE}) can be linearized in 
small fluctuations $\delta\phi(t,x)$ around the stationary 
classical background (\ref{GS}):
\begin{equation}
i\hbar\partial_t
\left(
\begin{array}{c}
\delta\phi   \\
\delta\phi^*
\end{array}
\right)~=~ 
{\cal L}~ 
\left(
\begin{array}{c}
\delta\phi   \\ 
\delta\phi^*
\end{array}
\right)~.
\nonumber
\end{equation}
Here the linear differential operator is
\begin{eqnarray}
&&{\cal L}=
\left(
\begin{array}{cc}
+ {\cal H}_{GP} + g|\phi_0(x)|^2 &
+ g\phi_0^2(x) \\
- g{\phi^*_0}^{2}(x) &
- {\cal H}_{GP} - g|\phi_0(x)|^2
\end{array}
\right)~.
\nonumber\\   
&&{\cal H}_{GP}=
-\frac{\hbar^2}{2m}\partial_x^2+g|\phi_0(x)|^2-\mu~,
\nonumber
\end{eqnarray}
The right eigenmodes of the non-hermitian ${\cal L}$ are solutions 
of the Bogoliubov-de Gennes equations
\begin{equation}
{\cal L}~
\left(
\begin{array}{c}
u \\
v
\end{array}
\right)~=~
\epsilon~
\left(
\begin{array}{c} 
u \\   
v
\end{array}
\right)~.
\label{BdG}
\end{equation}  
Every right eigenmode has a corresponding left eigenmode 
$(u^*,-v^*)$. The right eigenmodes can be classified as 
phonon modes ($\epsilon>0$) and zero modes ($\epsilon=0$).


\subsection{ Phonons }   

For later convenience I write first continuous spectrum $\epsilon_k$ 
and eigenfunctions for phonons on the background
$\phi_0(x)=e^{-i\theta}\sqrt{\rho}\tanh\frac{x-q}{\xi}$
extending to $x=\pm\infty$, 
\begin{eqnarray}
&& 
\epsilon_k=
\sqrt{\hbar^2c^2k^2+
      \left(\frac{\hbar^2k^2}{2m}\right)^2}~,\nonumber\\
&& 
u_k(x,q)= 
\frac{g\rho}{\sqrt{4\pi\xi}~\epsilon_k}
e^{ikx} 
e^{-i\theta}\times \nonumber\\
&&
\left[
\left(
(k\xi)^2+\frac{2\varepsilon_k}{g\rho}
\right)
\left(
\frac{k\xi}{2}+i\tanh\frac{x-q}{\xi}
\right)
+
\frac{k\xi}{\cosh^2\frac{x-q}{\xi}}
\right]~,
\nonumber \\
&& 
v_k(x,q)=
\frac{g\rho}{\sqrt{4\pi\xi}~\epsilon_k}
e^{ikx} 
e^{i\theta}\times \nonumber\\
&&
\left[
\left(
(k\xi)^2-\frac{2\varepsilon_k}{g\rho}
\right)
\left(
\frac{k\xi}{2}+i\tanh\frac{x-q}{\xi}
\right)
+
\frac{k\xi}{\cosh^2\frac{x-q}{\xi}}
\right]~.
\nonumber
\end{eqnarray}
The soliton does not scatter phonons but shifts their phase: as $u_k(x)$
or $v_k(x)$ is passing from left to right the function
$\left(\frac{k\xi}{2}+i\tanh\frac{x-q}{\xi}\right)$ is changing phase by
$\Delta \varphi_k=2\arctan\left(\frac{2}{k\xi}\right)$.


\subsection{ Phonons in a box }

In a box the wavevector $k$ is quantized. In figure \ref{figphi0} we can
see a half-antikink at $x=0$, a kink at $x=q$, and another half-antikink
at $x=l$. In this background total phaseshift of a phonon passing from the
left wall to the right wall of the box is
\change{$\frac12\Delta\varphi_k+\Delta\varphi_k+\frac12\Delta\varphi_k=0$} and the
quantization condition is
\change{$k_nl+2\Delta\varphi_{k_n}~=~n\pi$}. Discrete Bogoliubov modes are
\begin{eqnarray}
&&
\epsilon_n=\epsilon_{k_n}~,
\nonumber\\
&&
u_n(x)= 
\left\{
\begin{array}{lcl}
\change{           u_{-k_n}(-x, 0) +               u_{ k_n}(-x, 0)   } &,& \change{ x<x_L},     \\
\change{\alpha_n~  u_{ k_n}( x, q) +  \alpha_n^*   u_{-k_n}( x, q)   } &,& x_L<x<x_R, \\
\change{\alpha_n^2 u_{-k_n}(-x,-l) + \alpha_n^{*2} u_{ k_n}(-x,-l)   } &,& x_R<x,
\end{array}
\right. ,
\nonumber
\end{eqnarray}
\change{where $\alpha_n=-e^{i\Delta\varphi_{k_n}}$}, plus a formula for $v_n(x)$ obtained by replacing $u$ with $v$. The modes are $0$ at $x=0,l$. With proper normalisation phonon modes satisfy the orthogonality relation: $\langle u_m|u_n \rangle-\langle v_m|v_n \rangle=\delta_{mn}$.


\subsection{ Zero modes }

In addition to phonons there are two zero modes with $\epsilon=0$. One
originates \cite{CastinDum} from the global $U(1)$ gauge invariance $\phi
e^{-i\theta}\to\phi e^{-i(\theta+\epsilon)}$ broken by the classical
solution (\ref{GS}):
\begin{equation}
\left(
\begin{array}{c}
u_{\theta} \\
v_{\theta}
\end{array}
\right)=
i\hbar
\frac{\partial}{\partial\theta}
\left(
\begin{array}{c}
\phi_0 \\
\phi^*_0
\end{array}
\right)~,
\label{uvU1}  
\end{equation}
and the other from the translational invariance 
$q\to q+\epsilon$ broken by the solution (\ref{GS}):
\begin{equation}
\left(
\begin{array}{c}
u_q \\
v_q
\end{array}
\right)=
i\hbar
\frac{\partial}{\partial q}
\left(  
\begin{array}{c}
\phi_0 \\
\phi^*_0
\end{array}
\right)~.   
\label{uvq}
\end{equation}
The zero modes are orthogonal: $\langle u_\theta|u_q \rangle-\langle
v_\theta|v_q \rangle=0$.

Unlike phonons both zero frequency modes also have zero norms: 
$\langle u|u\rangle-\langle v|v \rangle=0$. As 
a result, phonon modes together with zero modes do not span the
whole Hilbert space in the functional space 
$(\delta\phi,\delta\phi^*)$. For example, to find a coordinate
of $(\delta\phi,\delta\phi^*)$ in the direction of the zero
mode $(u_\theta,v_\theta)$ one has to project $(\delta\phi,\delta\phi^*)$
on an {\it adjoint} vector $(u_{\theta}^{ad},v_{\theta}^{ad})$, 
\begin{equation}
\left( \langle u^{ad}_\theta|,-\langle v^{ad}_\theta| \right)
\left(\begin{array}{c}
|\delta\phi   \rangle \\
|\delta\phi^* \rangle
\end{array}\right)~.
\nonumber
\end{equation}
which has unit overlap with the zero mode, $\langle
u_{\theta}^{ad}|u_{\theta} \rangle- \langle v_{\theta}^{ad}|v_{\theta}
\rangle=1$, but is orthogonal to all other modes.

In a similar way, the translational mode $(u_q,v_q)$ requires an adjoint
mode $(u_q^{ad},v_q^{ad})$ normalized so that $\langle u_q^{ad}|u_q
\rangle- \langle v_q^{ad}|v_q \rangle=1$, but orthogonal to all other
modes including $(u_{\theta}^{ad},v_{\theta}^{ad})$. In summary, two
adjoint modes are missing in order to span the whole Hilbert space of
$(\delta\phi,\delta\phi^*)$.


\subsection{Adjoint modes}

In the case of the dark soliton (\ref{GS}) a vector $(u^{ad},v^{ad})$
adjoint to a zero mode $(u,v)$ turns out to be a solution of the
inhomogeneous equation
\cite{CastinDum}
\begin{equation} 
{\cal L}
\left(  
\begin{array}{c}
u^{ad} \\
v^{ad}
\end{array}
\right)=
\frac{1}{M}
\left(
\begin{array}{c}
u\\
v
\end{array}
\right)~   
\label{a}
\end{equation}
with a constant $M$ chosen so that the overlap $\langle u^{ad}|u
\rangle-\langle v^{ad}|v \rangle=1$. Eq.(\ref{a}) warrants that 
the adjoint vector $(u^{ad},v^{ad})$ is an eigenstate of 
${\cal L}^2$ with eigenvalue $0$. As such it is orthogonal to 
all phonon modes because phonon modes are eigenstates of
${\cal L}^2$ with non-zero eigenvalues $\epsilon_k^2$.

To find the adjoint vector to the gauge mode (\ref{uvU1}) it is
good to start form a stationary Gross-Pitaevskii equation solved
by $\phi_0$ in Eq.(\ref{GS}) 
\begin{equation}
0=-\frac{\hbar^2}{2m}\partial_x^2\phi_0+g|\phi_0^2|\phi_0-\mu\phi_0~.
\label{GPE}
\end{equation}
Taking derivative of this equation and its complex conjugate with respect 
to $\rho$ gives
\begin{equation}
{\cal L}
\left(
\begin{array}{c}
\partial_{\rho}\phi_0   \\
\partial_{\rho}\phi_0^*
\end{array}
\right)=
\frac{\partial\mu}{\partial\rho}
\left(
\begin{array}{c}
\phi_0   \\
-\phi_0^*
\end{array}
\right)~.
\nonumber
\end{equation}
Comparing this equation with (\ref{a}) and (\ref{uvU1}) and
using $\partial_{\rho}\mu=g$ gives
\begin{equation}
\left(
\begin{array}{c}
u^{ad}_{\theta} \\
v^{ad}_{\theta}
\end{array}
\right)=
\frac{1}{\hbar gM_{\theta}}
\left(
\begin{array}{c}
\partial_{\rho}\phi_0   \\
\partial_{\rho}\phi_0^*
\end{array}
\right)~.
\label{aU1}
\end{equation}
The normalization condition $\langle u^{ad}_{\theta}|u_{\theta} \rangle-
\langle v^{ad}_{\theta}|v_{\theta} \rangle=1$ requires that $M_{\theta}=
\frac{1}{g}\frac{\partial N_0}{\partial\rho}$. Here
$N_0=\langle\phi_0|\phi_0\rangle$ is average number of atoms in the
condensate mode. With this $M_{\theta}$ one recovers the general formula
\cite{CastinDum}
\begin{equation}
\left(
\begin{array}{c}
u^{ad}_{\theta} \\
v^{ad}_{\theta}
\end{array}
\right)=
\frac{\partial}{\partial N_0}
\left(
\begin{array}{c}
\phi_0   \\
\phi_0^*
\end{array}
\right)~.
\nonumber
\end{equation}  

To get the adjoint vector to the translational mode (\ref{uvq}) we  
verify first that
\begin{equation}
{\cal L}
\left(
\begin{array}{c}
e^{-i\theta}
\\
-e^{i\theta}
\end{array}
\right)
\frac{i\sqrt{\rho}}{c}
I(x)=
i\hbar
\frac{\partial}{\partial q}
\left(
\begin{array}{c}
\phi_0 \\
\phi^*_0
\end{array}
\right)~.
\end{equation}
Here the envelope function is
\begin{equation}
I(x)=
\left\{
\begin{array}{lcl}
\tanh\frac{x}{\xi}   &,& x<x_L      \\
1                    &,& x_L<x<x_R  \\
\tanh\frac{l-x}{\xi} &,& x_R<x
\end{array}
\right.
\end{equation}
Comparing this equation with Eqs.(\ref{a},\ref{uvq}) gives
\begin{eqnarray}
&&
\left(
\begin{array}{c}
u^{ad}_q \\
v^{ad}_q
\end{array}
\right)=
\frac{1}{M_q}  
\left(
\begin{array}{c}
e^{-i\theta} 
\\
-e^{i\theta}
\end{array}
\right)
\frac{i\sqrt{\rho}}{c}
I(x)
\label{at}
\end{eqnarray}
The normalization condition
$\langle u^{ad}_q|u_q \rangle-
\langle v^{ad}_q|v_q \rangle=1$ requires $M_q~=~-4\hbar\rho/c~<~0$. 

The overlap between two adjoint modes should vanish but it does not:
$\langle u^{ad}_q|u^{ad}_{\theta} \rangle- 
\langle v^{ad}_q|v^{ad}_{\theta} \rangle=iR$ with real 
$R=\frac{2q-l}{\hbar gc M_q M_\theta}$. 
This problem can be easily fixed because the solution
of the inhomogeneous equation (\ref{a}) is not unique - we can always add 
a zero mode of ${\cal L}$ to the solution. Using this freedom 
I replace the adjoint gauge mode as
\begin{eqnarray}
\left(      
\begin{array}{c}
u^{ad}_{\theta} \\
v^{ad}_{\theta}
\end{array}
\right)
&\to &
\left(      
\begin{array}{c}
u^{ad}_{\theta} \\
v^{ad}_{\theta}
\end{array}
\right)
-iR
\left(
\begin{array}{c}
u_q \\
v_q
\end{array}
\right)~.
\label{adthetato}
\end{eqnarray}
The new adjoint gauge mode has no overlap with the adjoint translational
mode.


\subsection{ Decomposition of unity }

Putting together phonons and zero modes with their
adjoint partners results in a decomposition of the unit operator as
\begin{eqnarray}
&&
\hat 1= \label{1}\\
&&
\left(\begin{array}{c}
|u_{\theta}\rangle \\
|v_{\theta}\rangle
\end{array}\right)
\left( \langle u^{ad}_{\theta}|,-\langle v^{ad}_{\theta}| \right)+
\left(\begin{array}{c}
|u^{ad}_{\theta}\rangle \\
|v^{ad}_{\theta}\rangle
\end{array}\right)
\left( \langle u_{\theta}|,-\langle v_{\theta}| \right)+
\nonumber\\
&&
\left(\begin{array}{c}
|u_q\rangle \\  
|v_q\rangle   
\end{array}\right)
\left( \langle u^{ad}_q|,-\langle v^{ad}_q| \right)+
\left(\begin{array}{c}
|u^{ad}_q\rangle \\
|v^{ad}_q\rangle
\end{array}\right)
\left( \langle u_q|,-\langle v_q| \right)+
\nonumber\\
&&
\sum_{n=1}^{\infty}~
\left(\begin{array}{c}
|u_n\rangle \\
|v_n\rangle
\end{array}\right)  
\left( \langle u_n|,-\langle v_n| \right)+
\left(\begin{array}{c}
|v_n^*\rangle \\
|u_n^*\rangle
\end{array}\right)
\left( \langle v_n^*|,-\langle u_n^*| \right)
\nonumber
\end{eqnarray}
This decomposition is a foundation of the Bogoliubov theory. 

\section{Perturbative Theory}

The Gross-Pitaevskii equation (\ref{TDGPE}) is a classical version
of a second quantized theory with a Hamiltonian
\begin{equation}
\hat H=
\int dx_0^l~
\left(
\frac{\hbar^2}{2m}
\partial_x\hat\psi^{\dagger}
\partial_x\hat\psi+
g\hat\psi^{\dagger}\hat\psi^{\dagger}\hat\psi\hat\psi
-\mu\hat\psi^{\dagger}\hat\psi
\right)~.
\label{hatH}
\end{equation}
Here $\hat\psi(x)$ is a bosonic field operator. 
In standard perturbative treatment \cite{CastinDum}
of zero modes the field operator
is expanded in small quantum fluctuations $\delta\hat\psi$ 
around the classical solution $\phi_0$ with fixed $q=q_0$ and $\theta=0$: 
$~\hat\psi=\phi_0+\delta\hat\psi$. The fluctuation operator
is expanded as
\begin{eqnarray}
\left(
\begin{array}{c}
\delta\hat\psi           \\
\delta\hat\psi^{\dagger}
\end{array}
\right)&=&
\sum_n~
\hat b_n
\left(
\begin{array}{c}
u_n \\
v_n
\end{array}
\right)_0+
\hat b_n^*
\left(
\begin{array}{c}
v_n^* \\
u_n^*
\end{array}
\right)_0+
\nonumber\\
&& 
\hat P_{\theta}
\left(
\begin{array}{c}
u^{ad}_{\theta} \\
v^{ad}_{\theta}
\end{array}
\right)_0+
\hat P_q
\left(
\begin{array}{c}
u^{ad}_q \\
v^{ad}_q
\end{array}
\right)_0+
\nonumber\\
&&
\frac{\hat\theta}{i\hbar}
\left(
\begin{array}{c}
u_{\theta} \\
v_{\theta}
\end{array}
\right)_0+
\frac{\hat q-q_0}{i\hbar}
\left(
\begin{array}{c}
u_q \\
v_q
\end{array}
\right)_0
\label{deltapsi}
\end{eqnarray}
Here the subscript $_0$ means a mode with $q=q_0$ and $\theta=0$.
Fluctuating position $\hat q-q_0$ and phase $\hat\theta$ are assumed
small. $\hat P_q$ and $\hat P_{\theta}$ are momenta conjugate to $\hat q$
and $\hat\theta$: $[\hat q,\hat P_q]=i\hbar$ and $[\hat \theta,\hat
P_{\theta}]=i\hbar$. $\hat b_n$ is a phonon annihilation operator:  
$[\hat b_m,\hat b_n^{\dagger}]=\delta_{mn}$. These commutation relations
plus the decomposition of unity (\ref{1}) give the desired
$[\delta\hat\psi(x),\delta\hat\psi^{\dagger}(y)]= \delta(x-y)$.

In the Bogoliubov theory the Hamiltonian (\ref{hatH}) is expanded to
second order in $\delta\hat\psi$. The linear term vanishes because the
classical $\phi_0$ is a solution of the Gross-Pitaevskii equation. The
leading second order term is a perturbative Bogoliubov Hamiltonian
\begin{eqnarray}
\hat H_{\rm pert.}&=&
\int dx 
\left( \delta\hat\psi^{\dagger},-\delta\hat\psi \right)
{\cal L}
\left(
\begin{array}{c}
\delta\hat\psi           \\
\delta\hat\psi^{\dagger}
\end{array}
\right)=
\nonumber\\
&&
\sum_n~
\epsilon_n \hat b_n^{\dagger} \hat b_n ~+~
\frac{\hat P^2_q}{2M_q} ~+~
\frac{\hat P^2_{\theta}}{2M_{\theta}}
\label{Hp}
\end{eqnarray}
As is well known \cite{CastinDum}, this theory is predicting
its own demise. The perturbative Hamiltonian (\ref{Hp}) predicts 
indefinite spreading of phase and position with time,
\begin{eqnarray}
\langle \hat\theta^2 \rangle ~\sim~ t~, 
\langle (\hat q-q_0)^2          \rangle ~\sim~ t~, 
\nonumber
\end{eqnarray}
while at the same time the derivation of (\ref{Hp}) requires
them to remain small. In the next Section
I introduce non-perturbative treatment of the zero modes 
that does not suffer from this inconsistency. 

\section{ Non-perturbative theory }

The key point is observation that any classical field can be 
expanded as
\begin{eqnarray}
\left(
\begin{array}{c}
\phi   \\
\phi^*
\end{array}
\right)&=&
\left(  
\begin{array}{c}  
\phi_0   \\        
\phi_0^*   
\end{array}
\right)+
\nonumber\\
&&
\gamma_{\theta}
\left(  
\begin{array}{c}
u^{ad}_{\theta} \\
v^{ad}_{\theta} 
\end{array}
\right)+
\gamma_q
\left( 
\begin{array}{c}
u^{ad}_q \\
v^{ad}_q
\end{array}
\right)+\nonumber\\
&&
\sum_n~
b_n
\left(
\begin{array}{c}
u_n \\
v_n
\end{array}
\right)+   
b_n^*
\left(
\begin{array}{c}
v_n^* \\
u_n^* 
\end{array}
\right)
\label{phiBog}
\end{eqnarray}
Here $b_n$'s are complex Bogoliubov amplitudes. For the lower component to
be a complex conjugate of the upper component the coordinates
$\gamma_{\theta}$ and $\gamma_q$ must be real, compare 
Eqs.(\ref{aU1},\ref{at},\ref{adthetato}). The real collective
coordinates $\theta$ and $q$ are implicit in definitions of
$\phi_0(x)$, adjoint modes and Bogoliubov modes. Unlike in the
perturbative treatment (\ref{deltapsi}), here $\theta$ and $q$ are not
assumed to be small, but they can take non-perturbatively large
values.

\subsection{ Effective Hamiltonian }

The Gross-Pitaevskii equation (\ref{TDGPE}) follows from a 
Lagrangian
\begin{eqnarray}
L=
\int_0^l dx~ 
\left(~
i\hbar\phi^*\partial_t\phi-
\frac{\hbar^2}{2m}|\partial_x\phi|^2-
g|\phi|^4+\mu|\phi|^2~
\right)
\nonumber
\end{eqnarray}
Substitution of Eq.(\ref{phiBog}) to $L$, expansion to second order in
$\gamma$ and $b$, and subsequent integration over $x$ give an effective
Lagrangian for the collective coordinates:
\begin{eqnarray}
L_{\rm eff}&=&
\sum_n~
\left(     
i\hbar b_n^* \dot b_n - \epsilon_n b_n^* b_n
\right)~+\nonumber\\
&&  
\hbar N \dot\theta ~+~
P \dot q~
-\frac{\gamma_{\theta}^2}{2M_{\theta}}
-\frac{\gamma_q^2}{2M_q}~.
\label{Leff}
\end{eqnarray}
Here $N$ is a total number of atoms (conjugate to $\theta$)
\begin{equation}
N=
\frac12 i
\left[
\langle \phi | \partial_{\theta} \phi \rangle -
{\rm c.c.}
\right]= 
N_0~+~
\frac{\gamma_{\theta}}{\hbar}+{\cal O}(\gamma\gamma,b\gamma,bb)~,
\end{equation}
and $P$ is center of mass momentum (conjugate to $q$)
\begin{equation}
P= 
\frac12 i\hbar
\left[
\langle \phi | \partial_q \phi \rangle-
{\rm c.c.}
\right]~=
\gamma_q+{\cal O}(\gamma\gamma,b\gamma,bb)~.
\end{equation} 
Legendre transformation leads to an effective Hamitonian
\begin{equation}
H_{\rm eff}~=~
\sum_n~
\epsilon_n b_n^* b_n
+\frac{\gamma_{\theta}^2}{2M_{\theta}}
+\frac{\gamma_q^2}{2M_q}~.
\nonumber
\end{equation}
To complete the transformation $\gamma$'s must be expressed 
as functions of the canonical momenta $N$ and $P$:
$
\gamma_{\theta}^2
\approx
\hbar^2 (N-N_0)^2~
$
and
$
\gamma_q^2
\approx
P^2
$.
These approximate $\gamma_{\theta}$ and $\gamma_{q}$ result
in a non-perturbative Bogoliubov Hamiltonian
\begin{eqnarray}
H_{\rm eff} & = &
\sum_n~
\epsilon_n b_n^* b_n +
\frac{\hbar^2(N-N_0)^2}{2M_{\theta}}+
\frac{P^2}{2M_q}~.
\end{eqnarray}

\subsection{ Quantum dark soliton }

$H_{\rm eff}$ can be quantized by replacing {\it c}-numbers with
operators. Non-zero commutators are: $[\hat b_k,\hat
b_p^{\dagger}]=\delta(k-p)$, $[\hat\theta,\hat N]=i$, and $[\hat q,\hat
P]=i\hbar$. With constant $N_0$ we also have $[\hat\theta,\hat N-N_0]=i$.  
A quantum Hamiltonian in coordinate representation where
$\hat N-N_0=-i\partial_{\theta}$ and $\hat P=-i\hbar\partial_q$ is
\begin{eqnarray}
\hat H_{\rm eff}&=&
\sum_n~
\epsilon_n\hat b_n^{\dagger}\hat b_n 
~-~\frac{\hbar^2}{2M_{\theta}}\partial_{\theta}^2
~-~\frac{\hbar^2}{2M_q}\partial_q^2 ~.
\label{Hnp}
\end{eqnarray}
This Hamiltonian is a sum of phonon terms, a phase diffusion term,
and a soliton diffusion term with negative mass $M_q<0$.

\begin{figure}
\includegraphics[width=8.5cm]{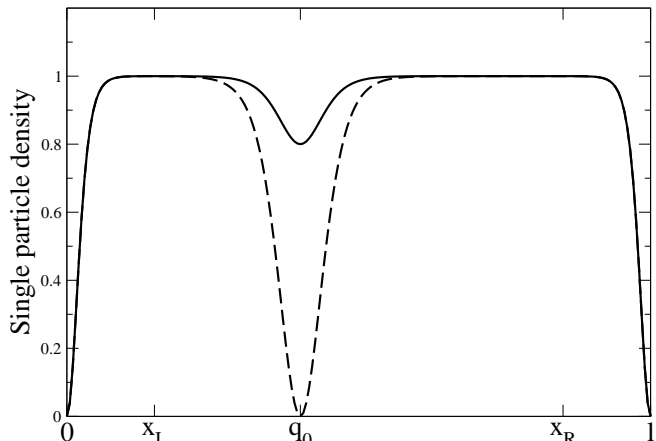}
\caption{
Quantum diffusion of soliton position. This figure shows the single 
particle density $\langle\hat\psi^{\dagger}(x)\hat\psi(x)\rangle/\rho$
at $t=0$ (dashed line) and at $t=\tau_{\rm diffusion}$ (solid line)
when the dispersion of soliton position $\Delta q$ equals
the soliton width $\xi$.
}
\label{figrho0}
\end{figure}

In the framework of the non-perturbative theory of zero modes one is free
to work in a subspace of Hilbert space with definite total number of atoms
$\hat N$. A state with definite $\hat N=N$ has a wave function
$\sim\exp(iN\theta)$ which covers the whole range of
$\theta\in[-\pi,+\pi)$ including non-perturbatively large values of
$\theta$. Such a wave function, and consequently also a state with
definite $\hat N$, is beyond reach for the perturbative theory. In the
subspace with $\hat N=N$ the Hamiltonian becomes
\begin{eqnarray}
\hat H_{\rm eff}^N&=&
\sum_n~
\epsilon_n\hat b_n^{\dagger}\hat b_n
~-~\frac{\hbar^2}{2M_q}\partial_q^2 ~.
\label{HnpN}
\end{eqnarray}
In the same way, a wave packet for soliton position $q$ can disperse
covering most of the box without contradicting assumptions of the theory.
For example, initial width $\Delta q_0$ of a gaussian wave packet grows
like
\begin{equation}
\Delta q(t)=
\sqrt{\Delta q_0^2+\frac{\hbar^2t^2}{4M_q^2~\Delta q_0^2}}
\stackrel{{\rm large }t}{\approx}
\frac{ct}{8~\rho~\Delta q_0}~.
\label{Deltaq}
\end{equation}
This dispersion of soliton position becomes comparable to the soliton
width, $\Delta q=\xi$, at the time $\tau_{\rm diffusion}=8\hbar\Delta
q_0/g$ which depends on the initial dispersion $\Delta q_0$.
In between $t=0$ and $t=\tau_{\rm diffusion}$ a hole in 
single particle density distribution fills up with atoms,
compare Fig.\ref{figrho0}.

It turns out that there is non-zero minimal uncertainty of soliton
position $\Delta q_{\rm min}>0$. Suppose that we have two condensates with
a dark soliton, but with different soliton positions $q_1$ and $q_2$. What
is the minimal distance between the solitons $|q_1-q_2|$ when it becomes
possible to distinguish these two condensates by a suitable quantum 
measurement? To answer this question we must calculate overlap
between these two condensates of $N$ atoms and see how it decays with the 
intersoliton distance,
\begin{eqnarray}
&&
\left(
\frac{\langle\phi_0^{q=q_1}|\phi_0^{q=q_2}\rangle}
     {\langle\phi_0^{q=\frac12(q_1+q2)}|\phi_0^{q=\frac12(q_1+q2)}\rangle}
\right)^N
\nonumber\\
&&
\stackrel{l\gg\xi}{\approx}
\left(
1-\frac{2(q_1-q_2)^2}{3l\xi}
\right)^N
\stackrel{l\gg\xi}{\approx}
\left(
1-\frac{2\rho(q_1-q_2)^2}{3N\xi}
\right)^N
\stackrel{N\gg 1}{\approx}
\nonumber\\
&&
\exp\left(-\frac{2\rho(q_1-q_2)^2}{3\xi}\right) ~.
\end{eqnarray}
Two condesates become orthogonal and in principle distinguishable when the
intersoliton distance becomes greater than $\Delta q_{\rm
min}=\sqrt{\frac{3\xi}{2\rho}}$. This is fundamental limitation derived
only from properties of the quantum states and not of any particular
measurement technique. This fundamental limitation means that the initial
gaussian wave packet cannot be localized better than $\Delta q_0=\Delta
q_{\rm min}$. This minimal dispersion leads to the minimal soliton 
diffusion time
\begin{equation}
\tau_{\rm diffusion}=
\frac{8\xi}{c}\sqrt{\frac{3\rho\xi}{2}}~.
\label{tau}
\end{equation}
It is interesting to evaluate $\tau_{\rm min}$ for the parameters of the
Hannover experiment \cite{Hannover}. This condensate can be approximated
by a quasi-1D harmonic trap like in Ref.\cite{greying}. In the present
paper we consider uniform condensate in a box with linear density $\rho$.
Using the linear density in the center of the effective quasi-1D trap as
$\rho$ we obtain $\tau_{\rm min}=8.0$ms. This is close to the time ${\cal
O}(10{\rm ms})$ when solitons are observed to loose contrast in that
experiment.

Finally, I make a brief comment on solitons moving with finite velocity
$v$ with respect to condensate. Condensate wave function for
$x\in\{x_L,x_R\}$ is $\phi_0^v=e^{-i\theta}\sqrt{\rho}
\left(i\beta+\alpha\tanh\alpha\frac{x-q-vt}{\xi}\right)$ with $\beta=v/c$
and $\alpha=\sqrt{1-\beta^2}$. Calculation of zero modes and adjoint
modes follows the same lines as for $v=0$. Mass of a moving soliton is
$M_q^v=\alpha M_q$ and the overlap between condensates decays when the
intersoliton distance becomes comparable to $\Delta q_{\rm min}^v=\Delta
q_{\rm min}/\alpha^{3/2}$. The minimal soliton diffusion time is
\begin{equation}
\tau_{\rm diffusion}^v=
\frac{\tau_{\rm diffusion}}{\sqrt{1-\frac{v^2}{c^2}}}~.
\label{tauv}
\end{equation}
For Hannover solitons \cite{Hannover} moving with velocities 
$v/c=0.4,\dots,0.8$
the minimal time is $\tau_{\rm min}^v=8.7\dots 13.3$ms.


\section{ Conclusion }

This paper develops non-perturbative theory of zero modes in the
Bogoliubov theory of atomic BEC. In the non-perturbative approach phase
fluctuations and fluctuations of soliton position are not restricted to
be small. This theory predics that an initially well localized soliton
wave packet is going to disperse beyond the soliton width.

When parameters of the present model are fit to the parameters of Hannover
experiment \cite{Hannover}, then dispersion of soliton position becomes
comparable to the soliton width after $10$ms from soliton creation. This
number is consistent with earlier perturbative studies of soliton
diffusion in a harmonic trap \cite{greying}. This diffusion time is also
consistent with the time when the dark soliton appears to loose contrast
or gray in Hannover experiment \cite{Hannover}.

Interaction of the soliton with a thermal cloud was offered as an
explanation \cite{Gora} of the observed graying. The quantum diffusion
described here and in Refs.\cite{greying} is a mechanism that operates
even in the $T=0$ quantum limit when the thermal cloud is turned off. The
quantum diffusion mechanism could be tested in an experiment with
variable temperature. Another dissipation mechanism due to non-uniform
condensate density was suggested in Ref.\cite{Durham}. The influence of
the inhomogenity could be elliminated in an experiment with a static
soliton ($v=0$) in the center of the trap. In principle it is possible to
make an experiment where the other mechanisms \cite{Gora,Durham} are
turned off leaving the quantum diffusion as the only known instability of
the dark soliton.

\section*{ Acknowledgements } 

I would like to thank Krzysztof Sacha for illuminating discussions, Grisha
Volovik for early contribution to this work, and Bogdan Damski for
comments on the manuscript. This research was supported in part by EU-IHP
programme (ULTI 3) and the KBN grant PBZ-MIN-008/P03/2003.

\change{I am indebted to George Zahariade for pointing out an error in Sec. III B. Corrections are marked in blue.}


\end{document}